\documentclass[preprint,jcp,nofootinbib]{revtex4}
\usepackage{amsfonts}
\usepackage{amsmath}
\usepackage{amssymb}
\usepackage{dcolumn}
\usepackage{epsfig}
\usepackage{subfigure}
\usepackage{bm}% bold math
\newcommand{\figurewidth}{4.0in}    % in preprint

\newcommand{\br}         {{\mathbf r}}
\newcommand{\bR}         {{\mathbf R}}
\newcommand{\bk}         {{\mathbf k}}
\begin{document}
\title{Complex coacervation: A field theoretic simulation study of polyelectrolyte complexation}
\author{Jonghoon Lee}
\email{jonglee@mpip-mainz.mpg.de} \altaffiliation[Present
address:]{Max Planck Institute for Polymer research, Ackermannweg
10, D-55128 Mainz, Germany}
\author{Yuri O. Popov}
\altaffiliation[Present address:]{Department of Physics, Case
Western Reserve University, Cleveland, Ohio 44106 }
\author{Glenn H. Fredrickson}
\altaffiliation[]{Department of Chemical Engineering and
Materials, University of California, Santa Barbara, California 93106}

\affiliation{Materials Research Laboratory, University of
California, Santa Barbara, California 93106-5121}
\date{\today}
%\received{\today}
%\revised{\today}
%\accepted{\today}
%\published{}

\begin{abstract}
Using the complex Langevin sampling strategy, field theoretic
simulations are performed to study the equilibrium phase behavior
and structure of symmetric polycation-polyanion mixtures without
salt in good solvents.  Static structure factors for the segment
density and charge density are calculated and used to study the role
of fluctuations in the electrostatic and chemical potential fields
beyond the random phase approximation. We specifically focus on the
role of charge density and molecular weight on the structure and
complexation behavior of polycation-polyanion solutions. A demixing
phase transition to form a ``complex coacervate'' is observed in
strongly charged systems, and the corresponding spinodal and binodal
boundaries of the phase diagram are investigated.
\end{abstract}
%\pacs{}
\maketitle

\section{Introduction}
%%polymer field theory and SCFT
Since statistical field theory was first applied to self avoiding
polymers by Edwards~\cite{Edw:1965}, polymer field theory models and
techniques have been developed and refined for a wide variety of
polymeric systems~\cite{deGennes:1969,Freed:1972,Helfand:1975,
Barrat:1996,Borukhov:1998,Netz:2003}.  In the field theoretic
approach, one often has to invoke approximations in order to make
functional integrals for partition functions and average properties
tractable. A convenient starting point is the mean field
approximation, also known as \textit{self-consistent field theory}
(SCFT)~\cite{Edw:1965,Helfand:1975}.  Since the effective
coordination number grows as the square root of the molecular weight
in concentrated polymeric systems, SCFT is argued to be
asymptotically exact in polymeric melts when the degree of
polymerization becomes infinite. Moreover, the theory serves as the
reference for more sophisticated approximation schemes that attempt
to account for field fluctuation effects~\cite{deGennes:1969,
deGennes:1979}. Over the last decade, with the ever increasing power
of digital computation, numerical SCFT has become a routine tool for
investigations of the structural and thermodynamic properties of
inhomogeneous polymers, including polymer alloys and block
copolymers of varying
architecture~\cite{Matsen:1996,Schmid:1998,Cochran:2006}.

%%However, correlation is everything for homogeneous charge system
Despite the success and the popularity of SCFT among polymer
researchers, there are classes of systems where SCFT is known to
be inaccurate~\cite{deGennes:1979}.  These systems, characterized
by strong or non-negligible density fluctuations, include polymer
solutions in dilute and semi-dilute regimes, systems close to a
phase transition, and polyelectrolyte solutions, among others. The
mean field approximation (i.e. SCFT) is particularly ill-suited to
polyelectrolytes; indeed, for some charged polymer systems the
Coulomb interaction does not contribute to the free energy at the
SCFT level. Such a system is a homogeneous polyelectrolyte
solution of arbitrary concentration in the bulk.  The overall
charge neutrality condition of the system leads to a constant
self-consistent electrostatic potential field that makes no
contribution to the free energy of the solution. Hence, in such a
system the electrostatics contributes to the free energy only via
positional \emph{correlations} among charges. Such charge
correlations can be very strong in polyelectrolytes due to the
high valency and the low translational entropy of macroions.

%Introduce Complex coacervation with applications
This strong charge correlation effect in polyelectrolytes was
noticed nearly a century ago~\cite{Bun:1929}: mixtures of
oppositely charged biopolymers or synthetic polyelectrolytes can
yield dense liquid precipitates coexisting with supernatant
solvent under standard physiological conditions at room
temperature.  This liquid-liquid phase separation, referred to as
\emph{complex coacervation}, is a manifestation of charge
correlations and fundamentally differs from the macrophase
separation that typically occurs in solutions or melts of
incompatible neutral polymers. Applications of complex coacervates
exist in both nature and technology. As an elegant example of the
former, the ``sand castle'' worm constructs habitats on the meter
length scale by gluing millimeter-sized grains of sand together
under the sea using a wonder glue composed of anionic and cationic
proteins~\cite{Zha:2005}. Technological applications span the
fields of water purification, adhesives, coatings, and
biotechnology. For example, DNA sensors are being developed based
on the complexation of target DNA molecules (anionic) with
synthetic conjugated cationic polyelectrolytes~\cite{Hon:2006}.
Other potential applications in biotechnology relate to drug
delivery and gene therapy, and invoke charge complexation to build
useful structures with a payload (such as DNA) for delivery in an
aqueous environment. An example of such a structure is the
self-assembled polymeric micelles produced by mixing solutions of
polyanions and polycations, where one or both polyelectrolytes
contains a charge neutral hydrophilic
block~\cite{Har:1999,Bur:2004,Kramarenko:2003,Kramarenko:2006}.
The core of each micelle is a coacervate composed of negatively
and positively charged polymer segments, while the corona consists
of water soluble neutral blocks that can serve to protect the
payload in the core.

%literature survey
Theoretically, the thermodynamics of complex coacervation was
first examined by Overbeek and Voorn~\cite{Mic:1957, Ove:1957}.
Although they correctly identified the driving force for the phase
separation as a Coulomb attraction between oppositely charged
macroions, they neglected the charge connectivity along the
polymer backbone and relied on the simple Debye-H\"{u}ckel
approximation in calculating the electrostatic free energy, which
is valid only in very dilute systems~\cite{Ove:1957}. Since that
time, the main issue in the theory of complex coacervation has
been how to properly calculate the electrostatic charge
correlation in polyelectrolyte solutions. Following the maturation
of polymer field theory, significant progress in our understanding
of charge correlation phenomena have resulted from application of
the random phase approximation (RPA) to weakly charged
polyelectrolytes~\cite{Borue:1988, Borue:1990, Kudlay:2004a,
Kudlay:2004b}, and loop expansions to treat stronger correlations
beyond the level of the RPA~\cite{Castelnovo:2001,
Castelnovo:2002, Oskolkov:2007}. Unfortunately these analytical
techniques are still limited in scope and are particularly
difficult to apply to \emph{inhomogeneous} structures, such as
mesophases.

% necessity of FTS
The pervasiveness and the importance of polyelectrolyte systems in
nature and biological applications, coupled with the obvious
inadequacy of SCFT in describing those systems, motivates the
development of a systematic \emph{numerical} way of incorporating
field fluctuation effects beyond SCFT. Ideally, such a scheme should
be capable of treating an arbitrary field theory model in the
absence of any approximations (aside from numerical errors that
arise from resolving and statistically sampling the fields), and
thus be capable of describing strong charge and density correlation
effects. Beyond serving as a general purpose simulation tool, such a
``field theoretic simulation'' method could be used as a test bed to
validate analytical results based on loop expansions and other
approximations.

% emerging FTS
Our group has developed a broad suite of methods for conducting
numerical simulations of polymer statistical field theory
models~\cite{Fre:2006}. Since the models typically have effective
Hamiltonians that are complex, rather than real, phase oscillations
thwart conventional numerical simulations based on Monte Carlo
sampling. We have found that this ``sign problem'' can be
effectively treated by adopting the complex Langevin (CL) sampling
method from nuclear physics~\cite{Par:1983,Kla:1983}, and applying
it in tandem with advanced numerical methods. This emerging field of
\emph{field theoretic simulations} (\emph{FTS}) of polymeric systems
has already been applied to several polymer models where SCFT is
known to be inaccurate, such as block copolymers near their
order-disorder transition (ODT)~\cite{Gan:2001,Fre:2002a}, ternary
blends close to a Lifshitz point~\cite{Due:2003}, and polymer
solutions in the semidilute regime~\cite{Ale:2003,Ale:2005}. The FTS
technique, however, is generic enough to be applicable to a much
broader class of field theory models of complex fluids, and is
potentially expandable to systems out of
equilibrium~\cite{Fre:2002b,Fre:2006}.

% outline of this article
In this article, we report on the application of FTS to
polyelectrolyte systems.  Based on a field theoretic model of a
simple binary polyelectrolyte mixture, FTS is used to study density
correlations in the system and to directly monitor the complexation
phenomena that leads to the formation of a coacervate. We recently
contributed a short highlight article that included some preliminary
findings~\cite{Yuri:2007}. Here we provide a more complete set of
results, a detailed discussion of our model and numerical methods,
and an in-depth analysis.  A particular focus of the present paper
is on charge and density correlations in the solution. Structure
factors of segment density and charge density are calculated using
FTS and compared with RPA structure factors.

\section{Theory and Numerical Methods}
In this section, the field theory of our model polyelectrolyte
solution is introduced. The thermodynamics of the system in the
mean field approximation is studied, followed by a discussion of
the first non-vanishing (RPA) correction to SCFT assuming small amplitude field
fluctuations. Next, the FTS technique is described along with CL
sampling.  The section ends with a discussion of the numerical
methods used to integrate the CL equations.

\subsection{Field theory model}
Here we propose a simple, yet fundamental, model that is capable
of giving rise to the phenomenon of complex coacervation with a
minimal set of parameters. Specifically, we consider a solution
consisting of a mixture of polycations and polyanions in an
implicit good solvent with a uniform dielectric constant
$\epsilon$. For simplicity, we choose to work with a
\emph{symmetric} model in which the two types of polyelectrolytes
are identical except for the sign of the charge that they carry.
Half of chain molecules in the system are positively charged and
the other half are negatively charged, thereby preserving the
condition of electroneutrality. Again, for simplicity, we do not
include either counterions or salt, although such generalizations
are straightforward. Our highly idealized model system, however,
could be approximately realized by mixing a polyacid with a
polybase of equal molecular weight and equal but opposite charge
in water. We formulate our model in the canonical ensemble and
consider a mixture of $n$ polyanions and $n$ polycations in a
volume $V$ of a three-dimensional physical space. The polymer
backbones are modeled as \emph{discrete} Gaussian (bead-spring)
chains with degree of polymerization $N$; the number of beads per
chain is $N+1$. In addition, we also employ the \emph{continuous}
Gaussian chain model to derive some analytical results in section
\ref{sec:IIC}; this model is a large-$N$ limit of the discrete
Gaussian chain. The charge per bead is $+z$ for the cation species
and $-z$ for the anion species; all other physical characteristics
of the two types of chains are assumed to be identical.
Interactions among beads include intramolecular spring forces, and
intra- and inter-molecular excluded volume and Coulomb
interactions.

The classical canonical partition function of the model just
described can be expressed (in ``particle'' form) as
\begin{equation}
\mathcal{Z}_{C}(2n, V, T) =
\frac{1}{{(n!)}^{2}{({{\lambda}_{T}}^{3})}^{2n(N+1)}}
\left(\prod_{\alpha=1}^{2n}\int d\bR_{\alpha}\right)
\exp[{-\beta{U_0}-\beta{\bar{U}_{ev}}-\beta{\bar{U}_{es}}}]\,,\label{eq:Zp}
\end{equation}
where ${\lambda}_{T}$ is the thermal de Broglie wave length for
unconnected beads and $\beta = 1/(k_B T)$ is the reciprocal of the
thermal energy. In indexing chains with $\alpha$, the first $n$
chains are the polyanions ($1\leq\alpha\leq n$) and the remaining
$n$ chains are the polycations ($n+1\leq\alpha\leq 2n$). The chain
conformation vector $\bR_{\alpha}$ is a set of coordinates of beads
belonging to chain $\alpha$: $\bR_{\alpha} = \{\br_{\alpha,0},
\br_{\alpha,1}, \br_{\alpha,2},\cdots, \br_{\alpha,N}\}$. The set of
bead coordinates for all $2n$ chains is denoted by $\bR$.
${\beta}U_0$ is the intramolecular, short-ranged potential that
connects beads along the chain backbone:
\begin{align}
\beta{U_0}[\bR ]=\beta\sum_{\alpha=1}^{2n}
\sum_{j=1}^{N}h(|\br_{\alpha,j}-\br_{\alpha,j-1}|) .
\end{align}
Gaussian discrete chains are assumed with harmonic spring potential
${\beta}h(r)=3r^{2}/2b^2$, where $b$ is the length of each
statistical segment.

The remaining contributions to the potential energy are potentials
of mean force (denoted by an over bar) because the solvent is
treated implicitly. ${\beta}\bar{U}_{ev}$ is the excluded volume
interaction of the system:
\begin{align}\label{eq:ev}
\beta{\bar{U}_{ev}}[\bR ] &= \frac{\beta}{2}\iint d^{3}\br
d^{3}\mathbf{r}'\,\hat{\rho}_{n}(\br)\bar{u}(|\br -
\mathbf{r}'|)\hat{\rho}_{n}(\mathbf{r}') .
\end{align}
Here the microscopic segment (bead) number density operator
$\hat{\rho}_{n}(\br)$ is defined as the sum of the microscopic
polyanion segment density $\hat{\rho}_{-}(\br)$ and the
microscopic polycation segment density $\hat{\rho}_{+}(\br)$,
where
\begin{equation}
\hat{\rho}_{-}(\br) = \sum_{\alpha=1}^{n}\sum_{j=0}^{N}\delta{(\br
- \br_{\alpha,j})}
\end{equation}
and \begin{equation} \hat{\rho}_{+}(\br) =
\sum_{\alpha=n+1}^{2n}\sum_{j=0}^{N}\delta{(\br - \br_{\alpha,j})}.
\end{equation}
We also adopt Edwards' simple delta function model for the volume
interaction~\cite{Edw:1965}:
\begin{align}
\beta\bar{u}(|\br |)= {u}_{0}\delta(\br)\,,
\end{align}
where $u_0$ is the excluded volume parameter.

The final contribution to the energy, ${\beta}\bar{U}_{es}$, is the
electrostatic interaction between charges in the system:
\begin{align}\label{eq:es}
\beta{\bar{U}}_{es}[\bR ] = \frac{1}{2}\iint d^{3}\br
d^{3}\mathbf{r}'\,\hat{\rho}_{c}(\br)\frac{l_{B}}{|\br -
\mathbf{r}'|}\hat{\rho}_{c}(\mathbf{r}')\,,
\end{align}
where $l_{B}\equiv {\beta}e^2/\epsilon$ is a constant Bjerrum length;
$e$ is the charge of a proton. Since every bead carries charge $+z$
or $-z$, the microscopic charge density is defined as
\begin{align}
\hat{\rho}_{c}(\br) = -z\hat{\rho}_{-}(\br)
+z\hat{\rho}_{+}(\br)\,.
\end{align}
By electroneutrality, the volume integral of this microscopic charge
density vanishes.

Our next step is to utilize Hubbard-Stratonovich transformation to
convert the ``particle'' representation of the partition function
in Eq.~(\ref{eq:Zp}) to a more convenient statistical field
theory. In this process, two auxiliary fields, $\phi$ and $w$, are
introduced and the partition function can be recast in the
form~\cite{Fre:2006}
\begin{equation}\label{eq:Zc}
\mathcal{Z}_C = \mathcal{Z}_0\int\mathcal{D}w\int\mathcal{D}\phi\,
\exp{\left(-H[w,\phi]\right)}\,.
\end{equation}
where the functional integrals over the two fields are taken in
the real function space and $H[w,\phi ]$ is a complex effective
Hamiltonian. The field $w(\br)$ can be interpreted as a
fluctuating chemical potential field conjugate to
$\hat{\rho}_{n}(\br)$, while $\phi(\br)$ is an electrostatic
potential field conjugate to $\hat{\rho}_{c}(\br)$.
$\mathcal{Z}_{0}$ is a prefactor that includes the ideal gas
entropy of non-interacting chains and some spurious
self-interactions contained in Eqs.~(\ref{eq:ev}) and
(\ref{eq:es}). These self-interactions produce only a constant
shift in chemical potential and have no thermodynamic consequence.
The effective Hamiltonian corresponds to the functional
\begin{equation}
H[w, \phi] = \frac{1}{2}\int d^{3}\br\,
\left(\frac{[w(\br)]^{2}}{u_0} + \frac{{|\nabla\phi(\br)
|}^{2}}{4\pi l_{B}}\right)\,-n\ln {Q[iw + iz\phi]}-n\ln {Q[iw -
iz\phi]}\,,\label{eq:H}
\end{equation}
where $Q[iw\pm iz\phi]$ is a single chain partition functional of
\emph{decoupled polyelectrolytes} in the conjugate fields.
Specifically, $Q$ is defined as the ratio of the partition function
of a single chain subject to the (pure imaginary) fields $iw(\br)\pm
iz\phi(\br)$ to the partition function of an ideal chain,
\begin{align}
Q[iw\pm iz\phi]= \frac{\int{d^{3}\br}^{N+1}\,\exp[-\beta
U_{s}(\br^{N+1})]}{V{\left(\int{d^{3}\mathbf{b}}\,\exp[-{\beta}h(|\mathbf{b}|)]\right)}^{N}}\,,
\end{align}
where $\br^{N+1}$ denotes the bead coordinates of the chain and
\begin{align}
\beta U_{s}(\br^{N+1})= \sum_{j=1}^{N}{\beta}h(|\br_{j}-\br_{j-1}|)
+ i\sum_{j=0}^{N}[w(\br_{j}){\pm}z\phi(\br_{j})] .
\end{align}
The functional $Q[iw\pm iz\phi ]$ is thus normalized so that
$Q[0]=1$. In practice, $Q$ is computed for an arbitrary field
$\psi(\br)$ according to $Q[\psi]=\frac{1}{V}\int d\br^{3}\,q(\br,
N;[\psi])$, where $q(\br, j;[\psi])$ represents the statistical
weight for the $j$th bead of a chain to be at position $\br$. This
object $q(\br,j;[\psi])$, referred to as a chain propagator, is
calculated by iterating the following Chapman-Kolmogorov type
equation
\begin{align}
q(\br,j+1;[\psi])={\left(\frac{3}{2\pi{b}^2}\right)}^{3/2}
\exp[-\psi(\br)]\int{d^{3}\br'}\,q(\br',j;[\psi])
\exp\left(-\frac{3|\br-\br'|^2}{2b^2}\right) \label{eq:CK}
\end{align}
from the initial condition of $q(\br,0;[\psi])=\exp[-\psi(\br)]$.
It is notable that the integral on the right hand side of
Eq.~(\ref{eq:CK}) is of convolution form so can be efficiently
evaluated using Fourier transforms.

The average of any thermodynamic observable $\mathcal{G}$ can be
formally defined as an ensemble average of the corresponding
operator $\tilde{\mathcal{G}}[w,\phi]$ over the auxiliary field
variables
\begin{equation}\label{eq:G}
\langle \mathcal{G}\rangle =
\frac{\mathcal{Z}_0}{\mathcal{Z}_C}\iint
\mathcal{D}w\mathcal{D}\phi \;
\tilde{\mathcal{G}}[w,\phi]\exp{\left(-H[w,\phi]\right)} .
\end{equation}
In the present paper, we are particularly interested in operators
for densities of polymer segments and charges, and operators whose
averages over the field variables yield two-point density and charge
correlation functions. The latter can be obtained by augmenting
Eq.~(\ref{eq:Zc}) with a source term involving external fields
conjugate to the microscopic segment and charge densities, and using
$\ln{\mathcal{Z}_{C}}$ as a generating functional for the cumulant
moments of density~\cite{Fre:2006}. The segment density operator for
polycations (polyanions) is found to be
\begin{equation}\label{eq:rhoQ}
\tilde{\rho}_{\pm}({\br};[w,\phi])\equiv -n\frac{\delta \ln Q[iw \pm
iz\phi]}{\delta (iw (\br ) \pm iz\phi (\br ))}\,,
\end{equation}
so that $\langle\hat{\rho}_{\pm}(\br)\rangle = \left
\langle\tilde{\rho}_{\pm}(\br;[w,\phi])\right \rangle$. In practice,
the functional derivative on the right hand side of Eq.~(\ref{eq:rhoQ})
can be computed using the chain propagator $q$ as
\begin{equation}
\frac{\delta \ln Q[\psi]}{\delta \psi (\br
)}=-\frac{\exp(\psi)}{VQ[\psi]}\sum_{j=0}^{N}
q(\br,N-j;[\psi])q(\br,j;[\psi])\,.
\end{equation}

The total segment number density operator $\tilde{\rho}_{n}(\br)$
is the sum of $\tilde{\rho}_{+}(\br)$ and $\tilde{\rho}_{-}(\br)$:
\begin{equation}\label{eq:rho}
\tilde{\rho}_{n}({\br};[w,\phi])= \tilde{\rho}_{+}({\br};[w,\phi])
+ \tilde{\rho}_{-}({\br};[w,\phi]) .
\end{equation}
Likewise, the charge density operator $\tilde{\rho}_{c}(\br)$ is
the sum of $\tilde{\rho}_{+}(\br)$ and $\tilde{\rho}_{-}(\br)$
weighted by the respective charge densities:
\begin{equation}\label{eq:rho_e}
\tilde{\rho}_{c}({\br};[w,\phi])=
z\left\{\tilde{\rho}_{+}({\br};[w,\phi]) -
\tilde{\rho}_{-}({\br};[w,\phi])\right\} .
\end{equation}

The correlations in the density fluctuations can be formally related
to the correlation functions of the corresponding auxiliary fields.
The pair correlation function of total segment number density
 can be computed from the pair correlation
function of the auxiliary chemical potential field according to
\begin{equation}
\langle\hat{\rho}_{n}(\br)\hat{\rho}_{n}(\br')\rangle=\frac{\delta(\br-\br')}{u_0}-\frac{\langle
w(\br)w(\br')\rangle}{{u_{0}}^{2}}\,. \label{eq:rhorho}
\end{equation}
The charge density correlation function can be similarly related to
the pair correlation function of the auxiliary electrostatic
potential field
\begin{equation}
\langle\hat{\rho}_{c}(\br)\hat{\rho}_{c}(\br')\rangle=\frac{-\nabla^{2}\delta(\br-\br')}{{4\pi}
l_{B}}-\frac{\langle
\nabla^{2}\phi(\br)\nabla^{2}\phi(\br')\rangle}{{({4\pi}
l_{B})}^{2}}\,.
\end{equation}
The cross correlation between the segment number density and the
charge density is similarly given by
\begin{equation}
\langle\hat{\rho}_{n}(\br)\hat{\rho}_{c}(\br')\rangle=\frac{\langle
w(\br)\nabla^{2}\phi(\br')\rangle}{{u_{0}}{4\pi} l_{B}}\,.
\end{equation}

The microstructure of the polyelectrolyte solution can be
characterized with static structure factors, which are related to
the pair correlation functions by Fourier transforms. The number
density structure factor, defined as
\begin{equation}
S_{nn}(k) =\frac{1}{V}\iint{d^{3}\br}{d^{3}\br'}\,
\exp{\left[-i{\bk}\cdot(\br-\br')\right]}
\langle\hat{\rho}_{n}(\br)\hat{\rho}_{n}(\br')\rangle \,,
\end{equation}
can be evaluated formally from Eq.~(\ref{eq:rhorho}) in terms of the
$w$ field pair correlation function
\begin{equation}\label{eq:SNN}
S_{nn}(k)
=\frac{1}{u_0}-\frac{\langle\hat{w}(\bk)\hat{w}(-\bk)\rangle}{{u_0}^{2}V}
\end{equation}
for $k\neq 0$, where $\hat{w}(\bk)$ is the Fourier transform of
$w(\br)$.  Likewise, the static structure factor for the charge
density can be calculated from the $\phi$ field pair correlation
function as
\begin{equation}\label{eq:Szz}
S_{cc}(k)=\frac{k^2}{4\pi l_{B}}- {\left(\frac{k^2}{4\pi
l_{B}}\right)}^2
\frac{\langle\hat{\phi}(\bk)\hat{\phi}(-\bk)\rangle}{V} \,,
\end{equation}
where $\hat{\phi}(\bk)$ is the Fourier transform of $\phi(\br)$.
The structure factor for the cross correlation between the segment
number density and the charge density can be calculated from the
correlation between the $w(\br)$ field fluctuation and the
$\phi(\br)$ field fluctuation,
\begin{equation}\label{eq:Snc}
S_{nc}(k)=- \frac{k^2}{u_{0}4\pi l_{B}}
\frac{\langle\hat{w}(\bk)\hat{\phi}(-\bk)\rangle}{V} \,.
\end{equation}

\subsection{Self consistent field theory}
Self-consistent field theory (SCFT) is derived by assuming that the
functional integrals of Eqs.~(\ref{eq:Zc}) and (\ref{eq:G}) are
dominated by ``mean field'' configurations $w^{*}(\br)$ and
$\phi^{*}(\br)$ that correspond to saddle points of the effective
Hamiltonian. The saddle point conditions produce the following SCFT
equations:
\begin{equation}
 \left.\frac{\delta{H}}{\delta w}
\right|_{{w=w^{*}}\atop{\phi=\phi^{*}} }=0
\stackrel{}{\longrightarrow}
\frac{w^{*}(\br)}{u_{0}}+i\tilde{\rho}_{n}({\br};[w^{*},\phi^{*}])
= 0\label{eq:mfw}
\end{equation}
and
\begin{equation}
\left.\frac{\delta{H}}{\delta\phi}
\right|_{{w=w^{*}}\atop{\phi=\phi^{*}} }=0
\stackrel{}{\longrightarrow}
\frac{-{\nabla}^{2}\phi^{*}(\br)}{4\pi
l_{B}}+i\tilde{\rho}_{c}({\br};[w^{*},\phi^{*}]) = 0
 .\label{eq:mfphi}
\end{equation}
The electrostatic mean field Eq.~(\ref{eq:mfphi}) recovers the
conventional Poisson equation with $ i\phi^*$ interpreted as the
electrostatic potential. Indeed, the physically relevant solutions
of these equations correspond to $w^*$ and $\phi^*$ being pure
imaginary fields. In an unbounded system, or a system with periodic
boundary conditions imposed on both $w(\br)$ and $\phi(\br)$, and
under good solvent conditions $u_0 >0$, these equations yield only
\emph{homogeneous} saddle fields of constant (imaginary) $w^{*}$ and
$\phi^{*}$. This trivial mean field solution corresponds to
$\nabla\phi^{*}=0$ and $w^{*}= -iu_{0}\rho_{0}$, where $\rho_{0}$ is
the average segment number density, $2n(N+1)/V$.

At the SCFT level of description, our model system is thus a
structureless polymer solution with constant segment density and a
trivial structure factor, $S_{nn}(k) = {u_{0}}^{-1}$. The overall
electro-neutrality makes the charge density vanish locally,
because the constant positive charge density field is compensated
by the constant negative charge density field. The Helmholtz free
energy $A$ thus involves only the ideal gas translational entropy
and the excluded volume interaction, and is minimal when polymer
chains are evenly distributed in the system:
\begin{align}
\beta{A}=
-\ln{\mathcal{Z}_C}&\approx-\ln{\mathcal{Z}_0}+H[w^{*},\phi^{*}]\nonumber\\
&= \beta{A_{0}}+\frac{u_{0}}{2}{\rho_{0}}^{2}V \,,
\end{align}
where $\beta{A_{0}}= -\ln{\mathcal{Z}_0}$ is the ideal chain
(translational entropy) term. Therefore, there is no contribution
of the Coulomb interaction to the mean field free energy, and SCFT
is unable to predict the formation of a complex coacervate.

\subsection{Gaussian fluctuations}\label{sec:IIC}
The phenomenon of coacervation can be traced to the presence of
charge correlations in polyelectrolyte mixtures -- correlations that
are neglected in the mean field approximation. This correlation
effect can be treated analytically with the systematic loop
expansion scheme. The first correction to SCFT in such a loop
expansion can be calculated analytically for our model system under
the assumptions of continuous (rather than discrete) Gaussian chains
and weak (low amplitude) field fluctuations. Later, the correlation
functions calculated in this subsection will be compared with FTS
results for discrete Gaussian chains with varying degrees of
polymerization $N$ and for arbitrary strengths of field
fluctuations.

For the case of a polymer that is experiencing an arbitrary field
$\psi(\br)$ that fluctuates only weakly from its homogeneous mean
field $\psi^*$, the single chain partition function $Q[\psi]$ for
a continuous Gaussian chain can be approximated up to the second
order in field fluctuation as~\cite{Fre:2006}
\begin{align}
Q[&\psi ]\approx \exp[-N\hat{\psi}(0)/V]\nonumber\\
\times &\left\{
1+\frac{N^2}{2V^2}\sum_{\bk\neq0}\hat{g}_{D}(k^{2}R_g^2)\hat{\psi}(\bk)
\hat{\psi}(-\bk) + \mathcal{O}({\hat{\psi}}^{3})\right\}
\,,\label{eq:Qapp}
\end{align}
where $\hat{\psi}(\bk)$ is the Fourier transform of $\psi(\br)$,
and $R_g=b\sqrt{N/6}$ is the radius of gyration of a Gaussian
chain without interactions. The Debye function
$\hat{g}_{D}({k^{2}R_g^2})$ is the scattering function of an ideal
continuous Gaussian chain~\cite{Doi:1986}:
\begin{equation}\label{eq:debye}
\hat{g}_{D}(x)=\frac{2}{x^2}\left[\exp{(-x)}+x-1\right] .
\end{equation}
With Eq.~(\ref{eq:Qapp}), the effective Hamiltonian of
Eq.~(\ref{eq:H}) can be approximated as
\begin{align}
&H[w, \phi]\approx H[w^{*}, \phi^{*}]\nonumber\\
&+\frac{1}{2V}\sum_{\bk\neq
0}\left[\hat{\gamma}_{ww}(k)\hat{w}(\bk)\hat{w}(-\bk)+
\hat{\gamma}_{\phi\phi}(k)\hat{\phi}(\bk)\hat{\phi}(-\bk)\right]
\,,\label{eq:Happ}
\end{align}
where fluctuations of $w$ and $\phi$ fields are decoupled at
quadratic order with expansion coefficients of
\begin{equation}
\hat{\gamma}_{ww}(k)=\frac{1}{u_0}+{\rho_{0}}N\hat{g}_{D}(k^{2}R_g^2)
\label{eq:gammaw}
\end{equation}
and
\begin{equation}
\hat{\gamma}_{\phi\phi}(k)=\frac{k^2}{4\pi
l_{B}}+{\rho_{0}}Nz^{2}\hat{g}_{D}(k^{2}R_g^2) .
\label{eq:gammaphi}
\end{equation}

Using Eqs.~(\ref{eq:Zc}) and (\ref{eq:Happ}), the osmotic pressure
$p$ of the solution with weak Gaussian field fluctuations can be
approximated as
\begin{align}
&\beta{p}
=-{\left(\frac{\partial{\beta{A}}}{\partial{V}}\right)}_{n,\beta}
={\left(\frac{\partial{\ln{\mathcal{Z}_{C}}}}{\partial{V}}\right)}_{n,\beta}\nonumber\\
&\approx \frac{\rho_{0}}{N} + \frac{{u_0}{\rho_{0}}^{2}}{2}
-\frac{1}{24\pi}\left[{\left(\frac{12{u_0}{\rho_{0}}}{b^2}\right)}^{3/2}
+ \frac{1}{\sqrt{2}}{\left(\frac{48\pi l_{B}{z}^{2}{\rho_0}}{b^2}\right)}^{3/4}
\right] \,,\label{eq:pi}
\end{align}
where the last negative term with square brackets is the
correction to the mean field osmotic pressure due to field
fluctuation effects. Eqn.~(\ref{eq:pi}) can be expressed in a
dimensionless form,
\begin{equation}
\beta{p}{R_g}^{3}\approx C+\frac{BC^2}{2}
-\frac{1}{24\pi}\left[(2BC)^{3/2} +
\frac{(2EC)^{3/4}}{\sqrt{2}}\right],\label{eq:pi2}
\end{equation}
with reduced (dimensionless) variables of
\begin{equation}\label{eq:BCE}
C=\frac{2n{R_g}^3}{V},\,\,B=\frac{{u_0}{N}^2}{{R_g}^3},\,\,
E=\frac{4\pi l_{B}{z}^{2}{N}^2}{{R_g}},
\end{equation}
where $C$ is a reduced chain concentration, $B$ is a reduced
excluded volume parameter, and $E$ is a reduced Bjerrum length.

The comparison between contributions from the mean field and the
correction due to field fluctuations in Eq.~(\ref{eq:pi2}) provides
a criterion that can be used to assess the importance of field
fluctuations. For the excluded volume interaction, the $w$ field
fluctuation can be regarded as weak correction when
\begin{equation}\label{eq:bc}
BC^{2} \gg (BC)^{3/2}\,.
\end{equation}
Therefore, in a three dimensional physical space, the mean field
result becomes asymptotically exact when $C/B \gg
1$~\footnote{While the condition $C/B \gg 1$ is commonly
identified as the definition of the dense regime for neutral
polymer solutions~\cite{Fre:2006,Ale:2003}, we note that the
precise account of the numerical prefactors in Eq.~(\ref{eq:pi2})
produces a slightly different condition: $C/B \gg 1/{18 \pi^2}
\approx 1/178$. The numerical factor $1/{18 \pi^2}$ lowers the
formal boundary between the dense and the semi-dilute regimes by
more than two orders of magnitude.  For this reason we call the
case of $C=1$ and $B=12$~(discussed below) as ``moderately dense"
rather than ``semi-dilute", as this case may be interpreted as
being on either side of the formal boundary between the dense and
the semi-dilute regimes depending on the exact condition
applied.}. In contrast, the electrostatic $\phi$ field fluctuation
contribution, namely the term scaling like $~- (EC)^{3/4}$, is of
paramount importance because there is no contribution from the
mean electrostatic field to the osmotic pressure. Thus,
electrostatic correlation effects can be neglected only when
$(EC)^{3/4}$ is small compared with all other contributions to the
osmotic pressure. Because the electrostatic mean-field term is
identically zero, a criterion similar to Eq.~(\ref{eq:bc}) cannot
be derived from Eq.~(\ref{eq:pi2}), and hence the range of
validity of the electrostatic term is unknown at the one-loop
(Gaussian) level. Evaluation of higher order terms in a loop
expansion would be required to clarify the range of validity of
this term.

It is important to note that Eq.~(\ref{eq:pi2}) predicts a
\emph{negative} (\emph{attractive}) contribution of electrostatic
correlations to the osmotic pressure. This  contribution, which is
similar to expressions derived previously using the
RPA~\cite{Borue:1988, Borue:1990, Kudlay:2004a, Kudlay:2004b,
Castelnovo:2001, Castelnovo:2002, Oskolkov:2007}, can drive
complexation of polyanions and polycations to produce a complex
coacervate phase. As will be discussed below, the polymer
concentration in the coacervate can be estimated by balancing the
repulsive excluded volume terms and the attractive electrostatic
correlation terms in Eq.~(\ref{eq:pi2}).

Using auxiliary field pair correlations calculated with the
quadratic Hamiltonian of Eq.~(\ref{eq:Happ}), structure factors of
the segment density $S_{nn}(k)$ and charge density $S_{cc}(k)$ can
be approximated by expressions involving the quadratic expansion
coefficients of Eq.~(\ref{eq:gammaw}) and
Eq.~(\ref{eq:gammaphi})~\cite{Fre:2006}:
\begin{equation}
S_{nn}(k)\approx\frac{1}{u_0}-\frac{1}{{u_0}^{2}\hat{\gamma}_{ww}(k)}\label{eq:SNN_RPA}
\end{equation}
and
\begin{equation}
S_{cc}(k)\approx\frac{k^2}{4\pi l_{B}} - {\left(\frac{k^2}{4\pi
l_{B}}\right)}^2 \frac{1}{\hat{\gamma}_{\phi\phi}(k)}
.\label{eq:Szz_RPA}
\end{equation}
These structure factors are commonly referred to as RPA structure
factors and are applicable in the weak field fluctuation limit of
Eq.~(\ref{eq:Happ}) where the harmonic fluctuations in the $w$ and
$\phi$ fields are decoupled. Indeed, the structure factor for the
cross correlation between the number density and the charge density
$S_{nc}(k)$ vanishes in this level of description, as the
approximated Hamiltonian of Eq.~(\ref{eq:Happ}) does not involve a
term proportional to $\hat{w}(\bk)\hat{\phi}(-\bk)$. Complex
coacervation is a phase transition that occurs when the charge
density correlations are strong enough to influence the segment
density distribution. Therefore, it would seem that structure
factors must be calculated beyond the RPA level to accurately
describe the complexation process~\cite{Castelnovo:2001}.

\subsection{Field theoretic simulations}
The analytic treatment of field fluctuation effects in the previous
subsection is based on the weak inhomogeneity approximation of
Eqs.~(\ref{eq:Qapp}) and (\ref{eq:Happ}). Here we turn to a direct
numerical approach (FTS) that avoids any assumption of weak field fluctuations.

The essence of FTS is to devise an efficient numerical strategy of
multi-dimensional integration whereby thermodynamic averages
defined in Eq.~(\ref{eq:G}) can be evaluated. The conventional
Monte Carlo importance sampling strategy is problematic for that
purpose. Although the functional integrals in Eq.~(\ref{eq:G}) are
over strictly real functions  $w(\br)$ and $\phi(\br)$, the
effective Hamiltonian $H[w,\phi ]$ is a complex functional.
Therefore, the probability weight proportional to
$\exp{(-H[w,\phi])}$ is not positive definite unless the sampling
trajectory happens to be along a constant phase path of
$\exp{(-H[w,\phi])}$. Because the identification of constant phase
paths in high dimensional function spaces is computationally
unfeasible, a fundamentally different simulation technique is
required.

The complex Langevin (CL) sampling strategy addresses the so-called
``sign problem'' associated with the non-positive definite
statistical weight of the field theory~\cite{Par:1983, Kla:1983}.
The idea behind this method is to extend the real fields into the
complex plane and to compute ensemble averages of observable
quantities by sampling fields along a stationary stochastic
trajectory in the complex function space. By extending the real
fields $w(\br)$ and $\phi(\br)$ into the complex functions
$W(\br)=w(\br)+iw_{I}(\br)$ and
$\Phi(\br)=\phi(\br)+i\phi_{I}(\br)$, the ensemble average of Eq.
(\ref{eq:G}) can be reexpressed as
\begin{equation}\label{eq:ACL}
\langle \mathcal{G}\rangle = \iiiint
\mathcal{D}w\mathcal{D}w_{I}\mathcal{D}\phi\mathcal{D}\phi_{I} \;
\tilde{\mathcal{G}}[W,\Phi]P[w, w_I , \phi , \phi_I ],
\end{equation}
where $P[w, w_I , \phi , \phi_I ]$ is a \emph{real non-negative}
statistical distribution of fields replacing the complex weight of
Eq.~(\ref{eq:G}). This comes at the cost of doubling the number of
the configurational degrees of freedom~\cite{Fre:2006}:
\begin{equation}
\frac{\mathcal{Z}_{0}\mathrm{e}^{-H[w',{\phi}']}}{\mathcal{Z}_{C}} =
\iiiint\mathcal{D}w\mathcal{D}w_{I}\mathcal{D}\phi\mathcal{D}\phi_{I}
\, \delta[w'-w- i w_{I}]\delta[{\phi}'-\phi- i \phi_{I}] P[w, w_I ,
\phi , \phi_I ], \label{eq:P}
\end{equation}
where $w'$ and $\phi'$ are real fields. Although necessary and
sufficient conditions for the existence of $P$ satisfying
Eq.~(\ref{eq:P}) given a complex effective Hamiltonian $H$ have been
identified~\cite{Sal:1997, Wei:2002}, these conditions are difficult
to verify in the highly nonlinear and nonlocal Hamiltonians of
polymer field theory. However, the current and the previous
successful applications of the CL method to polymer models provide
such a proof \textit{a
posteriori}~\cite{Ale:2003,Ale:2005,Gau:1993}.

The CL dynamics is a stochastic Langevin dynamics in the complex
function space designed to generate a stationary Markov sequence of
complex functions with distribution $P[w, w_I , \phi , \phi_I ]$:
\begin{align}
&\frac{\partial}{\partial{t}}W(\br ,t) =
-\lambda_{w}\left[\frac{\delta H[W,\Phi]}{\delta W(\br ,t)}\right] +
{\eta}_{w}(\br, t)\nonumber\\
&= -\lambda_{w}\left[
\frac{W(\br,t)}{u_{0}}+i\tilde{\rho}_{n}(\br,t;[W,\Phi]) \right] +
{\eta}_{w}(\br, t)\label{eq:CLW}
\end{align} and
\begin{align}
&\frac{\partial}{\partial{t}}\Phi(\br ,t)=
-\lambda_{\phi}\left[\frac{\delta H[W,\Phi]}{\delta {\Phi}(\br
,t)}\right] + {\eta}_{\phi}(\br, t) \nonumber\\
&= -\lambda_{\phi}\left[ \frac{-{\nabla}^{2}\Phi(\br,t)}{4\pi
l_{B}}+i\tilde{\rho}_{c}({\br,t};[W,\Phi]) \right] +
{\eta}_{\phi}(\br, t) ,\label{eq:CLPhi}
\end{align}
where $\eta_{w}(\br, t)$ and $\eta_{\phi}(\br, t)$ are \emph{real}
Gaussian white noise fields with zero mean and variances
proportional to the real dissipative coefficients $\lambda_{w}$ and
$\lambda_{\phi}$, respectively, consistent with the familiar
fluctuation-dissipation theorem of Brownian
dynamics~\cite{Fre:2006}. The CL equations should be interpreted as
a \emph{fictitious}, rather than physical, dynamics to sample the
field configuration space. In the absence of the random forces, the
above equations reduce to deterministic equations that have saddle
point field configurations of Eq.~(\ref{eq:mfw}) and
Eq.~(\ref{eq:mfphi}) as steady state solutions. With the forcing
terms, the stochastic dynamics drive trajectories in the complex
function space that produce steady distributions $P[w,w_I ,\phi ,
\phi_I ]$ that are peaked at saddle point configurations. Because
the random forces are strictly real, the imaginary components of the
CL dynamics drive the field trajectories towards constant phase
paths, while the real components of the equations are responsible
for stochastic motion along a path. Under conditions where a
stationary distribution is achieved, the ensemble average of
Eq.~(\ref{eq:ACL}) can be approximated by a time average along the
CL trajectory.

In the numerical application of CL dynamics to affect a
field-theoretic simulation, physical space and time are both
discretized. For bulk simulations with periodic boundary conditions
imposed on the fields, we have found that the spatial discretization
is most conveniently accomplished by spectral collocation using a
plane wave basis and a uniform computational grid of $M$ sites
\cite{gottlieb77}. Fast Fourier transforms (FFTs) can then be used
to efficiently switch between real space (i.e. an $M$-vector of
field values on the grid sites) and Fourier space (i.e. the first
$M$ Fourier coefficients) representations of the fields.

Upon spectral collocation, the continuum CL Eqs.~(\ref{eq:CLW}) and
(\ref{eq:CLPhi}) are transformed into a set of $4M$ stochastic
differential equations that can be integrated forward in time from
an initial field configuration by standard
algorithms~\cite{kloeden92,ottinger96}. The simplest algorithm is
the explicit Euler-Maruyama time integration scheme,
\begin{equation}
\bm{W}^{(t+{\Delta{t}})}=\bm{W}^{(t)}- \lambda_{w}\Delta{t}
{(\Delta{x})^3}\left(\frac{\bm{W}^{(t)}}{u_0} +
i{\bm{\rho}_{n}}^{(t)}\right)+ {\bm{R}_{w}}^{(t)} \label{euler1}
\end{equation}
and
\begin{equation}
\bm{\Phi}^{(t+{\Delta{t}})}= \bm{\Phi}^{(t)} -
\lambda_{\phi}\Delta{t}
{(\Delta{x})^3}\left(\frac{-\nabla^{2}\bm{\Phi}^{(t)}}{4\pi l_{B}}
+ i{\bm{\rho}_{c}}^{(t)}\right)+ {\bm{R}_{\phi}}^{(t)} ,
\label{euler2}
\end{equation}
where $\Delta{x}$ is the grid spacing used for the spacial
discretization, and $\Delta{t}$ is the time spacing used for the
temporal discretization. Here, $\bm{W}, \bm{\Phi}, \bm{\rho}_{n},$
and $\bm{\rho}_{c}$ are $M$-vectors of complex variables, which
represent the collocated values of the fields $W(\br), {\Phi}(\br),
\tilde{\rho}_{n}(\br),$ and $\tilde{\rho}_{c}(\br)$ on the
computational grid. The superscripts index the discrete time at
which the $M$-vectors are evaluated. ${\bm{R}_{w}}^{(t)}$ and
${\bm{R}_{\phi}}^{(t)}$ are $M$-vectors of real Gaussian random
variables, which are obtained by spatial collocation of the
continuous functions $R_{w}(\br,t)$ and $R_{\phi}(\br,t)$ defined by
\begin{equation}
{R_{w}(\br,t)} \equiv \int_{t}^{t+{\Delta{t}}}d\tau\,\eta_{w}(\br,
\tau)
\end{equation}
and
\begin{equation}
{R_{\phi}(\br,t)} \equiv
\int_{t}^{t+{\Delta{t}}}d\tau\,\eta_{\phi}(\br, \tau).
\end{equation}
The grid-collocated fields $\bm{R}_{w}$ and $\bm{R}_{\phi}$ are
Gaussian white noise functions with vanishing averages, $\langle{
{\bm{R}_{w}}^{(t)}}\rangle = \langle{{\bm{R}_{\phi}}^{(t)}}\rangle =
0$, second moments given by
\begin{equation}
 \langle \bm{R}_w^{(t)} \bm{R}_w^{(t^\prime )} \rangle =
2\lambda_{w}\Delta{t}\delta_{t,t^\prime} \bm{1}
\end{equation}
\begin{equation}
 \langle \bm{R}_\phi^{(t)} \bm{R}_\phi^{(t^\prime )} \rangle =
2\lambda_{\phi}\Delta{t}\delta_{t,t^\prime} \bm{1}
\end{equation}
and vanishing cross-correlations. The rank $M$ unit tensor is
denoted by $\bm{1}$.

Stochastic time integration algorithms for the CL equations lead
to time discretization errors in computed expectation values that
scale as a power of the time step $\Delta t$. The Euler-Maruyama
(EM) scheme summarized by Eqs.~(\ref{euler1}) and (\ref{euler2})
is of weak order one, which implies that the errors in computed
averages vanish as $\Delta t$ for small $\Delta t$. This low order
accuracy and the poor stability characteristics of the
Euler-Maruyama algorithm make it unsuitable for the large-scale 3d
simulations reported here. We have instead adopted a
semi-implicit, weak second order algorithm developed by Lennon and
coworkers~\cite{Len:pre}, which was itself inspired by operator
splitting methods devised by Petersen and
\"{O}ttinger~\cite{Pet:1998,ottinger96}. Beyond the second order
accuracy, the Lennon algorithm utilizes analytic information about
the linearized force in a semi-implicit update scheme to improve
stability. The algorithm can be cast in a predictor-corrector form
with the field updates conducted in Fourier space (discrete
Fourier transforms of the collocated fields are denoted by
carets). The predictor steps are explicit EM updates of
Eq.~(\ref{euler1}) and Eq.~(\ref{euler2}) in Fourier space:
\begin{equation}
{\hat{\bm{W}}}^{(*)} = {\hat{\bm{W}}}^{(t)} -
\Delta_{w}\left(\frac{{\hat{\bm{W}}}^{(t)}}{u_0} +
i{\hat{\bm{\rho}}_{n}}^{(t)}\right)+
{\hat{\bm{R}}_{w}}^{(t)}\label{eq:halfW}
\end{equation}
and
\begin{equation}
\hat{\bm{\Phi}}^{(*)}= \hat{\bm{\Phi}}^{(t)} -
\Delta_{\phi}\left(\frac{k^{2}\hat{\bm{\Phi}}^{(t)}}{4\pi l_{B}} +
i\hat{\bm{\rho}}_{c}^{(t)}\right)+ {\hat{\bm{R}}_{\phi}}^{(t)}\,,
\label{eq:halfPhi}
\end{equation}
where the parameters $\Delta_{w}$ and $\Delta{\phi}$ are defined
as $\Delta_{w}\equiv \lambda_{w}\Delta{t}{(\Delta{x})}^{3}$ and
$\Delta_{\phi}\equiv \lambda_{\phi}\Delta{t}{(\Delta{x})}^{3}$.
The corrector steps are
\begin{equation}
{\hat{\bm{W}}}^{(t+\Delta{t})}= \frac{ {\hat{\bm{W}}}^{(t)}
+\left[1+ \Delta_{w}{\rho_{0}}N\hat{g}_{D}(k^{2}R_g^2) \right]
{\hat{\bm{W}}}^{(*)}  - \Delta_{w}i{\hat{\bm{\rho}_{n}}}^{(*)} +
{\hat{\bm{R}}_{w}}^{(t)}}
{2+\Delta_{w}{\hat{\gamma}_{ww}}(k)}\label{eq:Wssi}
\end{equation}
and
\begin{equation}
{\hat{\bm{\Phi}}}^{(t+\Delta{t})}= \frac{ {\hat{\bm{\Phi}}}^{(t)}
+\left[1+\Delta_{\phi}{\rho_{0}}N{z}^{2}\hat{g}_{D}(k^{2}R_g^2)\right]{\hat{\bm{\Phi}}}^{(*)}
- \Delta_{\phi}i{\hat{\bm{\rho}_{c}}}^{(*)}+
{\hat{\bm{R}}_{\phi}}^{(t)}}{2+\Delta_{\phi}{\hat{\gamma}_{\phi\phi}}(k)}\,,\label{eq:Phissi}
\end{equation}
where ${\hat{\bm{\rho}_{n}}}^{(*)}$ and
${\hat{\bm{\rho}_{c}}}^{(*)}$ are the segment number density and the
charge density operators based on the predicted fields of
$\hat{\bm{\Phi}}^{(*)}$ and $\hat{\bm{W}}^{(*)}$.

This improved stochastic integration algorithm was recently applied
to FTS-CL simulations of block copolymer melts and has been
extensively tested in that context by Lennon and
coworkers~\cite{Len:pre}. In the present case of polyelectrolyte
solutions, we have found that the Lennon algorithm permits the use
of a time step that is an order of magnitude larger than that
mandated by stability for the EM algorithm. This translates to a
ten-fold reduction in computational time.

In the following section, the FTS results are discussed in the
context of the dimensionless parameters introduced in
Eq.~(\ref{eq:BCE}): $B$, $C$ and $E$. These parameters appear
naturally in the field theory if all lengths are scaled by the
radius of gyration $R_g$, the $W$ field is rescaled according to
$\tilde{W} = N W$, and the $\Phi$ field is rescaled as $\tilde{\Phi}
= z N \Phi$. With these scalings, the CL equations
(\ref{eq:halfW})-(\ref{eq:Phissi}) for \emph{continuous} polymer
chains depend on the three intensive model parameters $(B, \; C, \;
E)$, the dimensionless simulation cell size $\tilde{L} = L/R_g$, and
on three dimensionless parameters that relate to the spatial and
temporal resolution of the numerical algorithm:
\begin{equation}
\widetilde{\Delta x} = \Delta x/ R_g
\end{equation}

\begin{equation}
(\Delta t)_w = \lambda_w N^2 \Delta t
\end{equation}

\begin{equation}
(\Delta t)_\phi = \lambda_\phi N^2 z^2 \Delta t
\end{equation}

For \emph{discrete} polymer chains, the number of segments $N$
appears as an additional independent parameter in the update
equations. However, we shall see that its influence is diminished as
$N$ is increased to large values approaching the continuous polymer
limit.

At fixed spatial resolution with the Lennon algorithm, we expect
second-order convergence in average properties as the parameters
$(\Delta t)_w$ and $(\Delta t)_\phi$ are reduced. All simulation
data reported below were obtained by setting $(\Delta t)_w$ and
$(\Delta t)_\phi$ to values such that the time integration error was
negligible compared to the statistical sampling error. In contrast,
due to ultraviolet divergences in the continuum field
theory~\cite{Fre:2006,Ale:2005}, we do not expect convergence of
certain average properties (such as absolute chemical potentials) as
the spatial grid spacing $\widetilde{\Delta x}$ is refined. However,
the structure factors and phase boundaries studied here were found
to be devoid of ultraviolet divergences, so no special
regularization procedures were required to isolate the
singularities.

\section{Results and Discussion}

In this section, we summarize and discuss FTS results for the field
theory model formulated in the previous section in the parameter
space of $B$, $C$, and $E$. While these three parameters completely
determine the intensive thermodynamic properties of a system
comprised of continuous Gaussian chains, as discussed above, there
is an additional independent parameter for the discrete Gaussian
chains employed in the simulations -- the degree of polymerization
$N$. We begin with homogeneous solutions of \emph{neutral} polymers
corresponding to $E=0$. The effect of the strength of the excluded
volume interaction ($B$) and the degree of polymerization ($N$) on
segment density correlations are studied using FTS, and compared
with the RPA structure factor. While studying the effect of $N$, the
parameters $C$ and $B$ are maintained at constant values. Next, we
discuss the effect of the parameter $E$ on density correlations and
charge correlations in oppositely charged polyelectrolyte solutions
with fixed $B$, $C$, and $N$. Finally, we construct the phase
diagram of our model system in a restricted region of the parameter
space. Although a more comprehensive study of the phase diagram is
worthy to pursue, the high dimensionality of the parameter space and
the significant computational requirements both limit the scope of
our investigation. Nonetheless, our results shall serve to highlight
the power and capability of FTS in addressing difficult problems in
polyelectrolyte complexation.

The cubic simulation box applied in our simulations has a volume
$V$ of ${(4R_{g})}^{3}$ ($\tilde{L} =4$) and is subject to
periodic boundary conditions. Occasionally, a larger simulation
box with $V$ of ${(8R_{g})}^{3}$ was used to confirm that finite
size effects were not influencing structure or thermodynamics. In
discretizing the simulation volume, $\Delta{x}$ is desired to be
smaller than the physical length scales of interest (i.e. the
correlation lengths for the two fields), but an excessively small
$\Delta x$ adds significantly to the computational cost. In most
of our simulations, $\Delta{x}$ was chosen to be $R_{g}/8$, so the
volume of ${(4R_{g})}^{3}$ corresponds to a lattice with $M= 32^3$
sites. Unlike $\Delta{x}$, which was fixed, $\Delta{t}$ was varied
over a broad range to achieve a consistent accuracy of time
integration depending on the strength of field fluctuations around
the mean field: $(\Delta t)_{w,\phi} \in [0.001,0.1]$. The
correlation time is also highly variable throughout parameter
space. Typically, $10^4$ statistically independent field
configurations were sampled to calculate averages, which
corresponds to a number of time steps in the range of
$5\times10^4$ to $5\times10^5$. The calculation of each
structure-factor curve reported here takes from $4$ days to $7$
weeks of CPU time on a single AMD Opteron 248 processor
($2.2$GHz). However, the algorithm scales nearly linearly with
domain decomposition across multiple processors, so the
simulations reported here are ideally suited for a parallel
computing environment.

\subsection{The segment density correlation in neutral polymer solutions}
By setting the charge on the polymer segments to zero ($z=E=0$), our
model system becomes a solution of electrically neutral polymers in
a good solvent. This is the Edwards' model~\cite{Edw:1965} (Model A
of Ref.~\cite{Fre:2006}). When the $w$ field fluctuations are weak,
by application of Eqs.~(\ref{eq:gammaw}) and (\ref{eq:BCE}), the
segment structure factor of Eq.~(\ref{eq:SNN_RPA}) can be
approximated by the dimensionless form,
\begin{align}\label{eq:SNN_BC}
u_{0}S_{nn}(k) \approx \frac{BC
\hat{g}_{D}(k^{2}{R_g}^{2})}{1 + BC \hat{g}_{D}(k^{2}{R_g}^{2})},
\end{align}
which indicates that the structure factor of segment density is
completely determined by one parameter, the product of $B$ and $C$.
In deriving this analytical formula, the chain molecules were
assumed to be continuous Gaussian chains (CGCs), thus the Debye
function $\hat{g}_{D}(k^{2}{R_g}^{2})$ depends only on the
unperturbed radius of gyration $R_g$. However, our simulations were
conducted using \emph{discrete} Gaussian chains (DGCs) with $N$
segments on each chain, for which the corresponding Debye functions
depend on $N$ as well as $R_g$. Therefore, the corresponding RPA
structure factor for DGCs can be written as
\begin{equation}\label{eq:SNN_BCN}
u_{0}S_{nn}(k, N) \approx \frac{BC
\hat{g}_{D}(k^{2}{R_g}^{2}, N)}{1 + BC \hat{g}_{D}(k^{2}{R_g}^{2},
N)},
\end{equation}
where~\cite{Doi:1986}
\begin{equation}\label{eq:debyeN}
\hat{g}_{D}(k^{2}{R_g}^{2},
N)=\frac{1}{{(N+1)}^{2}}\sum_{i,j=0}^{N}\exp{\left[-{(k
R_g)}^{2}\frac{\left|i-j\right|}{N}\right]}.
\end{equation}

\begin{figure}[tb]
\includegraphics[clip=true,angle=0,width=\figurewidth]{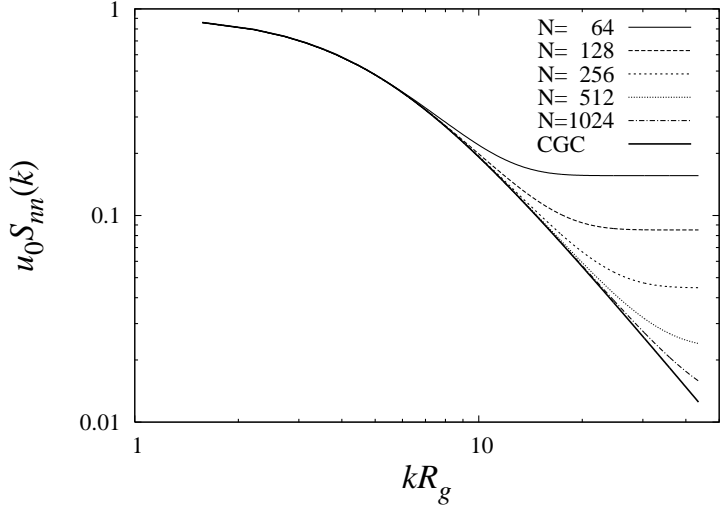}
\caption{RPA structure factor for the segment density when $BC=12$.
Segment structure factors for several values of $N$, calculated from
Eq.~(\ref{eq:SNN_BCN}), are compared with the result based on the
continuous Gaussian chain (CGC), cf.
Eq.~(\ref{eq:SNN_BC}).}\label{fig:SNN_RPA}
\end{figure}

RPA segment structure factors for DGCs of various $N$ are compared
with that of the CGC in FIG.~\ref{fig:SNN_RPA}. At fixed $BC$, the
low $k$ behavior of the structure factor, related to the isothermal
osmotic compressibility, is independent of the level of
discretization of the constituent chain molecules. It is only in the
high $k$ regime (compared with $2\pi/R_g$) where the discrete nature
of the DGC model manifests itself by saturating the $1/k^2$ decay of
the structure factor.

%comparison with FTS
\begin{figure}[tb]
\centering \mbox {
    \subfigure[Dense solution: $B=1$ and $C=12$]
    {
        \includegraphics[clip=true,angle=0,width=\figurewidth]{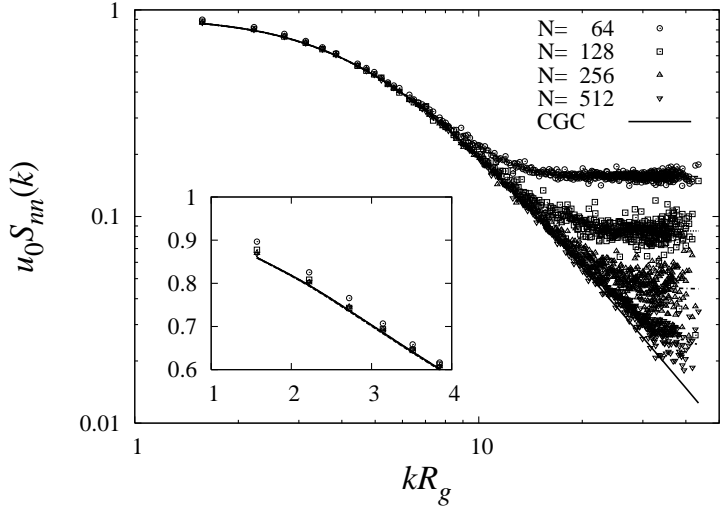}
            }
} \mbox {
    \subfigure[Moderately-dense solution: $B=12$ and $C=1$]
    {
        \includegraphics[clip=true,angle=0,width=\figurewidth]{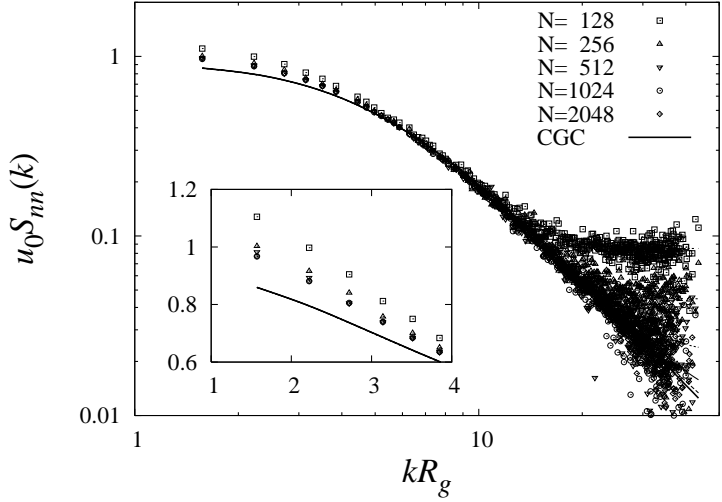}
    }
} \caption{The segment structure factor in solutions of neutral
discrete Gaussian chains: Symbols are FTS results calculated from
Eq.~(\ref{eq:SNN}). Lines are RPA structure factors from
FIG.~\ref{fig:SNN_RPA}. A three dimensional lattice of $M=32^3$
sites was employed with periodic boundary conditions. The cell
volume was $V=(4 R_g )^{3}$. $u_0$ was varied as $N$ was changed to
keep $B$ fixed at either $1$ or $12$. }\label{fig:SNN_N}
\end{figure}

As expected, the RPA structure factor proves to be a valid
approximation to $S_{nn} (k)$ as long as the $w$ field fluctuations
are weak and approximately Gaussian. This has been verified by
comparing the RPA structure factor to results obtained from FTS,
which incorporates the full field fluctuation spectrum. In
FIG.~\ref{fig:SNN_N}, structure factors derived from FTS for two
 different solution conditions are compared with the RPA
structure factor. FIG.~\ref{fig:SNN_N}(a) corresponds to the case
of a dense solution ($C > B$) with $B=1$ and $C=12$, while
FIG.~\ref{fig:SNN_N}(b) describes a moderately-dense solution with
$B=12$ and $C=1$. However, both share the same RPA predictions of
FIG.~\ref{fig:SNN_RPA}, because the RPA segment structure factor
only depends on the product $BC$ which is fixed at $12$. As
FIG.~\ref{fig:SNN_N}(a) shows, there is very good quantitative
agreement between RPA structure factors and FTS-derived structure
factors when the solution is dense ($C/B=12$). However, in the
moderately-dense regime, as exemplified by
FIG.~\ref{fig:SNN_N}(b), the RPA breaks down, especially at low
$k$. Due to strong excluded volume correlations in this regime,
the system evidently has a larger osmotic compressibility than is
predicted by the RPA.

Additionally, the FTS results at low $k$ show that the strength of
the $w$ field fluctuations actually depends on $N$. As implied in
Eq.~(\ref{eq:pi2}), the $w$ field fluctuations tend to increase
the osmotic compressibility of the system, and the FTS results
indicate that this tendency gets stronger for smaller $N$ (see
insets of low $k$ regime in FIG.~\ref{fig:SNN_N}). While a
discrete chain with $N\gtrsim 256$ is sufficient to model
$N$-independent thermodynamic properties in a dense solution with
$C/B=12$, a substantially larger $N$ ($\gtrsim 1024$) is required
for $N$-independent thermodynamics in a moderately-dense solution
of $B/C=12$.

\begin{figure}[tb]
\includegraphics[clip=true,angle=0,width=\figurewidth]{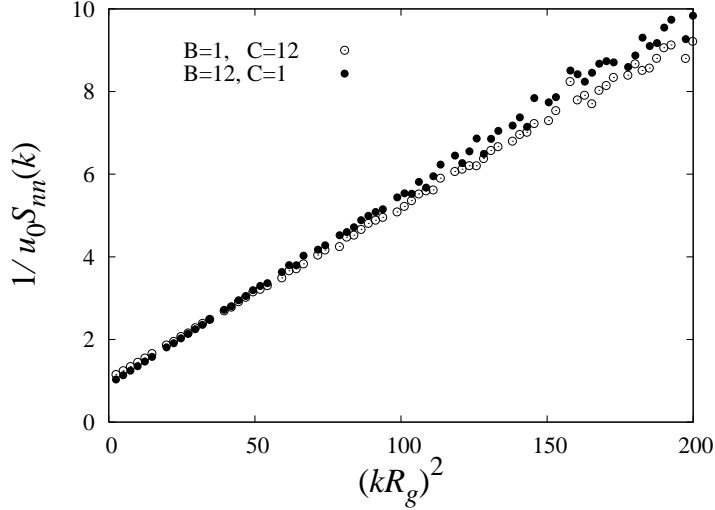}
\caption{Inverse structure factor plot used to extract the
correlation length for segment density, $\xi_u$. According to
Eq.~(\ref{eq:xi_u}), $\xi_u$ can be estimated from the ratio of
the slope to intercept in the intermediate $k$ region of this
plot, $1/R_{g}< k/2\pi< 1/\xi_{u}$. We find that $\xi_{u}/R_g =
0.201\pm 0.0014$ in the case of $B=1$ and $C=12$, and $\xi_{u}/R_g
= 0.225\pm 0.0016$ for $B=12$ and $C=1$. The Edwards correlation
length is $\xi_{E}/R_g \simeq 0.204$ for $BC=12$. The simulations
were conducted on a $M=32^3$ lattice with periodic boundary
conditions, a system volume of $V=(4 R_g )^{3}$, and a chain
length of $N=1024$. The linear fit was applied in the regime of
${(kR_g)}^2\in[30, 200]$. }\label{fig:xi_u}
\end{figure}

At larger $k$ and for large $N$, we can apply an asymptotic
expression for the Debye function, $\hat{g}_{D}(k^{2}
{R_g}^2)\approx 2/k^2 {R_g}^2$, which is highly accurate for $kR_g>
2\pi$. Using this approximation, the RPA structure factor for the
segment density (Eq.~(\ref{eq:SNN_BC})) can be rearranged as
\begin{equation}\label{eq:xi_u}
\frac{1}{u_{0} S_{nn}(k)}\approx 1+{\xi_{u}}^{2}{k}^{2} ,
\end{equation}
where $\xi_u$ is the correlation length for segment density. We
use Eq.~(\ref{eq:xi_u}) to define $\xi_u$ even outside of the RPA,
but note that the RPA predicts that $\xi_u = \xi_E$, where
$\xi_{E}=R_g/{\left(2BC\right)}^{1/2}$ is the Edwards correlation
length.  In FIG.~\ref{fig:xi_u}, segment correlation lengths are
estimated from FTS structure factors, which quantitatively agree
with a prior FTS study that used an independent method of
estimating $\xi_u$~\cite{Ale:2003}. Not surprisingly, the Edwards
correlation length is a good approximation to the segment
correlation length extracted from FTS when the solution is dense.
The segment correlation length in a moderately-dense solution,
however, is underestimated by the Edwards correlation length.

\subsection{Correlations in symmetric polycation-polyanion mixtures}

In this subsection, density correlations in homogeneous
polyelectrolyte solutions are studied with FTS. Unlike the results
presented so far for uncharged polymer solutions, charge
correlations, as well as  segment correlations, are of interest in
polyelectrolytes. Specifically, we are interested in the interplay
between charge correlations and segment correlations and how they
relate to the phenomenon of complex coacervation. In all the FTS
results reported in this subsection, the solution was dense
($C=12$, $B=1$) and discrete Gaussian chains of $N=1024$ were
used.

The RPA structure factor of Eq.~(\ref{eq:SNN_BC}) for the segment
density is evidently independent of the charge content in the
system; this is a result of decoupling of $w$ and $\phi$ fluctuations at the RPA level. It would seem that without a
higher-order analysis of fluctuations, it is impossible to guess
even the qualitative effect of charge correlations on segment
correlations. However, the one loop result for the osmotic pressure
does provide some insight. From Eq.~(\ref{eq:pi2}), the charge
correlations are seen to reduce the osmotic pressure of a
homogeneous solution, the same effect caused by the segment
correlations. Thus, by adding equal and opposite charges to the
chains of a neutral polymer solution, we can expect to obtain a
solution with increased osmotic compressibility. This increase in
compressibility will in turn be manifest in the low $k$ behavior of
$S_{nn}(k)$.

\begin{figure}[tb]
\includegraphics[clip=true,angle=0,width=\figurewidth]{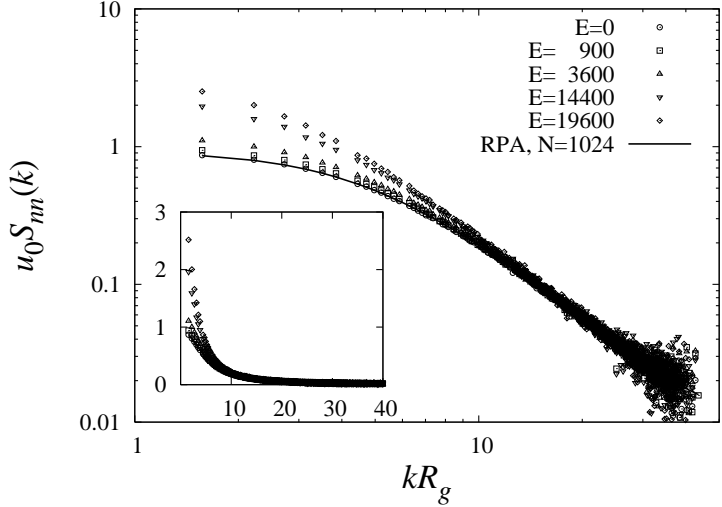}
\caption{The segment structure factor in solutions of symmetric
polyelectrolytes. The symbols are FTS results and the line is
calculated from the RPA result, Eq.~(\ref{eq:SNN_BCN}). The
following parameters were used: $B=1$, $C=12$, $N=1024$, $V=4\times
4\times 4 {R_g}^{3}$, and $M=32\times 32\times
32$.}\label{fig:SNN_E}
\end{figure}

In FIG.~\ref{fig:SNN_E}, segment structure factors for symmetric
polyelectrolyte solutions were obtained from FTS. Because the
solution is dense, the RPA structure factor is observed to be a
good approximation when the polymers are neutral ($E=0$). However,
as the charge density is increased from zero, the FTS results show
strong deviations from the RPA structure factor which is
independent of $E$. The osmotic compressibility increases
monotonically with $E$ and eventually diverges at even higher $E$,
indicative of a macrophase separation (in this case ``complex
coacervation'').

Another structure factor of interest in the charged system relates
to the charge density correlation.  When the $\phi$ field
fluctuations are weak, the RPA formula for the charge density
structure factor, Eq.~(\ref{eq:Szz_RPA}), can be combined with
Eqs.~(\ref{eq:gammaphi}) and (\ref{eq:BCE}) to obtain the following
approximation:
\begin{align}\label{eq:Szz_CE}
\frac{4\pi l_{B}}{k^2}S_{cc}(k) \approx \frac{CE
\hat{g}_{D}(k^{2}{R_g}^{2})}{k^2 R_g^2 + CE
\hat{g}_{D}(k^{2}{R_g}^{2})}.
\end{align}
This RPA formula also applies to solutions of discrete Gaussian
chains when the modified Debye function of Eq.~(\ref{eq:debyeN}) is
substituted for Eq.~(\ref{eq:debye}). We see from
Eq.~(\ref{eq:Szz_CE}) that the RPA structure factor for the charge
density is completely dictated by the combination parameter $CE$,
and is independent of $B$.

By again utilizing the asymptotic expression for the Debye
function, $\hat{g}_{D}(k^{2} {R_g}^2)\approx 2/k^2 {R_g}^2$, valid
for $N \gg 1$ and $kR_g> 2\pi$, the inversion of
Eq.~(\ref{eq:Szz_CE}) provides
\begin{equation}\label{eq:xi_z}
\frac{k^2}{4\pi l_{B} S_{cc}(k)}
 \approx 1+{\xi_{c}}^{4}{k}^{4} ,
\end{equation}
which defines a new charge correlation (or screening) length
$\xi_{c}$. Explicit use of the RPA formula leads to the result
$\xi_{c} = \xi_{PE}$, where $\xi_{PE} =
R_g/{\left(2EC\right)}^{1/4}$.

Eqn~(\ref{eq:xi_z}) also predicts that $S_{cc}(k)$ is
\emph{maximal} at $k=1/\xi_{c}$. Note that the correlation length
$\xi_{PE}$ is proportional to the $-1/4$ power of the segment
density, and hence is qualitatively different from the
Debye-H\"{u}ckel length for small ions, which is proportional to
the $-1/2$ power of the ion density. Thus, the attachment of
charges to polymer chains creates a coupling between chain
conformational statistics and charge density that drastically
changes the electrostatic correlation properties of
polyelectrolyte solutions compared with a conventional small-ion
electrolyte.

\begin{figure}[tb]
\includegraphics[clip=true,angle=0,width=\figurewidth]{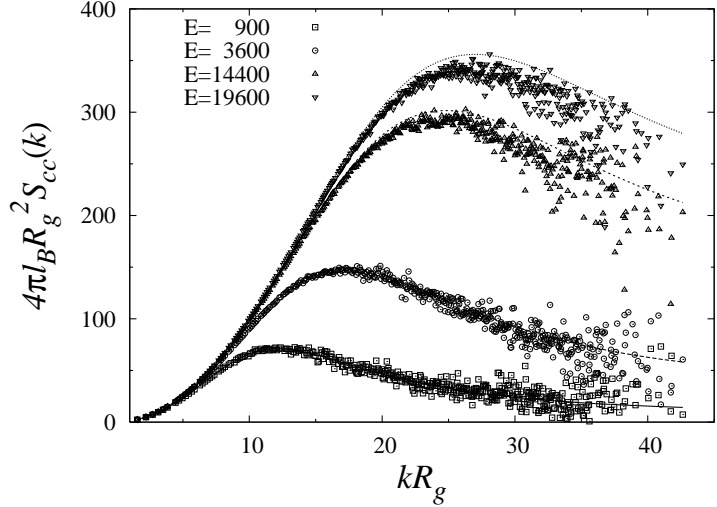}
\caption{The charge structure factor in solutions of oppositely
charged polyelectrolytes. Symbols are FTS results, while the lines
are RPA structure factors calculated from Eq.~(\ref{eq:Szz_CE}). In
the simulations, we vary the parameter $E$ by changing $z$ while
keeping $l_B$ constant. The parameters correspond to: $l_{B}=R_g/8$,
$B=1$, $C=12$, $N=1024$, $V=(4 R_g )^3$, and
$M=32^3$.}\label{fig:Szz_E}
\end{figure}

 In FIG.~\ref{fig:Szz_E}, FTS results for the charge
density structure factor are compared with RPA predictions based
on Eq.~(\ref{eq:Szz_CE}). The dimensionless object $4\pi
l_{B}{R_{g}}^{2}S_{cc}$, instead of $4\pi l_{B}S_{cc}/k^2$, is
plotted to clearly show the location of the maximum ($\sim
1/{\xi}_{c}$) and the screening of charge density at low $k$:
$S_{cc}(k)\sim k^{2}$. Although the RPA remains a good
approximation at low charge content ($E=900, \; 3600$), the
deviation between the RPA and FTS becomes noticeable at larger $E$
and larger $k$. As $E$ increases, the maxima of both the RPA and
FTS results shift to higher $k$, indicating a shorter screening
length consistent with the RPA scaling $\xi_{c}\sim (EC)^{-1/4}$.
At $E=19600$, however, the RPA structure factor places the maximum
at a somewhat higher value of $k$ than FTS. In other words, the
RPA underestimates the charge correlation length for strongly
charged chains. It is also evident that the RPA slightly
overestimates the amplitude of charge correlations at large $E$.
Later, it will be shown that this qualitative observation is
consistent with differences observed between phase boundaries
deduced from FTS studies and the one-loop analysis.

\subsection{Complex coacervation: phase diagram}

As implied by the diverging osmotic compressibility in
FIG.~\ref{fig:SNN_E}, a macrophase separation (complex
coacervation) is possible in our field theory model. It was
already anticipated in the analytic one-loop correction to the
osmotic pressure, Eq.~(\ref{eq:pi2}), that a phase separation may
result from the competition between the positive second virial
term from the excluded volume interaction and the negative
contribution from charge correlations. The spinodal (single-phase
stability limit) is determined by the condition $\partial \beta
p/\partial C = 0$ from the one loop osmotic pressure expression of
Eq.~(\ref{eq:pi2}). With an additional approximation regarding the
dilute phase, the one-loop theory can also predict the binodal
(the coexistence curve of dilute and coacervate phases). Assuming
a supernatant dilute phase free of polyelectrolyte, we can write
the binodal equation in the form of $\beta p=0$.

\begin{figure}[tb]
\includegraphics[clip=true,angle=0,width=\figurewidth]{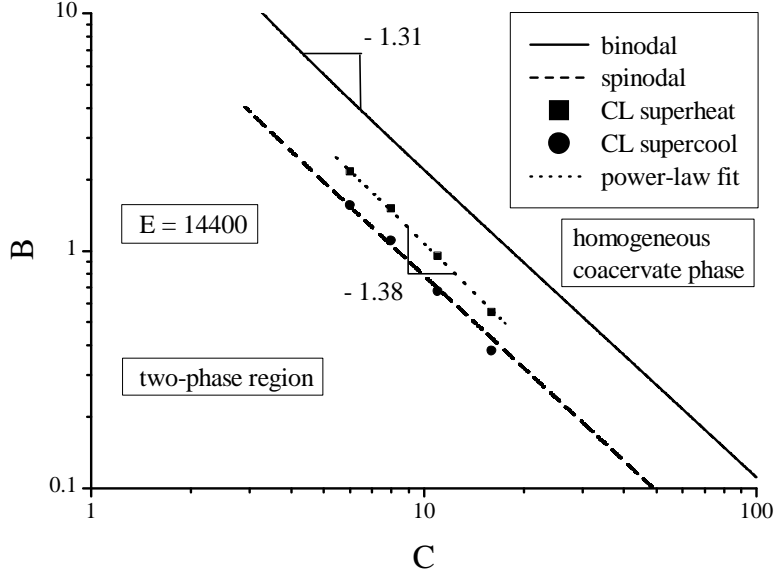}
\caption{ Phase diagram for the three-dimensional symmetric
polyelectrolyte mixture expressed in the coordinates of reduced
polymer concentration $C$ and reduced excluded volume $B$ at fixed
reduced Bjerrum length $E=14400$. Solid and dashed lines are the
analytical one-loop binodal and spinodal, respectively. The
one-phase region is above and to the right of the lines, and the
two-phase region is below and to the left. Symbols are from the
examination of the hysteresis in $\Delta{\rho}$ using FTS; details
of the procedure are explained in the text. FTS data obtained from
supercooling simulations (lowering $B$) are denoted by $\bullet$,
while superheating data are denoted by $\blacksquare$. The dotted
line is a power-law fit. The error in the numerical data is
comparable to the size of the symbols. Numerical simulations were
conducted in a cubic cell of size $(4 R_g )^3$ with periodic
boundary conditions and with $N=384$.}\label{phasediagram}
\end{figure}

In the context of computer simulations, a rigorous construction of
a phase diagram requires a computational technique for accessing
the free energy of the system. In the case of particle-based
simulations, a variety of free energy estimation methods are
available, including thermodynamic integration techniques, Gibbs
ensemble and particle insertion methods, and histogram
techniques~\cite{allen89,frenkel01}. Such methods, however, are
only now emerging for field-based simulations, so here we make a
crude estimate of the boundary of the two-phase region in our
polyelectrolyte model by monitoring the hysteresis of an order
parameter while varying an intensive parameter of the system. The
order parameter chosen to characterize complex coacervation is the
density difference $\Delta \rho$ between the dense coacervate
phase and the dilute phase. Hysteresis is examined while varying
the solvent quality $B$. For example, consider a homogeneous
single phase solution ($\Delta{\rho}=0$) at certain values of $B$,
$C$, and $E$. While slowly ``cooling'' that system by gradually
decreasing $B$ at fixed $C$ and $E$, a sudden jump in
$\Delta{\rho}$ from zero to a finite value occurs as the system
separates into two phases of different densities. The high density
phase can be identified as the complex coacervate. On the other
hand, upon ``heating'' the system from the two-phase region by
gradually increasing $B$ at fixed $C$ and $E$, a sudden drop of
$\Delta{\rho}$ from a finite value to zero is observed as the
system exceeds the limit of superheating and remixes into one
homogeneous phase. For most phase transitions, the phase
coexistence curve is closer to the superheating curve than to the
supercooling curve. Thus, while the superheating and supercooling
curves should bracket the transition, we expect that the
superheating curve will lie closer to the binodal boundary.

An example of a phase diagram (in 3D) constructed in such a way is
provided in Fig.~\ref{phasediagram}.  Spinodals and binodals are
surfaces in the three-parameter space of the reduced variables
$C$, $B$, and $E$. The figure represents a cross-section of this
three-dimensional space by a plane $E = 14400$;  hence, the
diagram involves only the $C$ and $B$ variables.  The one-phase
region (disordered homogeneous phase) is above and to the right of
the lines, and the two-phase region is below and to the left.  The
tie lines in the two-phase region are horizontal (constant $B$)
and connect nearly pure solvent ($C\approx 0$) with a coacervate
phase at the binodal concentration. Remarkably, the analytical and
numerical results nearly coincide in the high concentration region
of the figure, despite the limitations of each method.  The
numerical results are subject to finite cell size and chain
discretization limitations ($N=384$), while the analytical
predictions neglect two-loop and higher order terms in the
fluctuation analysis. Nonetheless, our numerical supercooling
result practically follows the analytical spinodal, and the
analytical binodal yields similar exponent ($-1.31$) as is
obtained from a power-law fit to the numerical superheating points
($-1.38$). The over-estimate of the size of the two phase region
by the one-loop theory is consistent with the observation in
FIG.~\ref{fig:Szz_E} that the RPA structure factor over-estimates
the strength of charge correlations at large $E$ and hence expands
the two-phase region. The overall semi-quantitative agreement
between theory and FTS, however, indicates that both approaches
have utility for this class of problems.

\section{Conclusions and Perspective}

In this paper, we reported on the application of the emerging
field-theoretic computer simulation (FTS) technique to a simple
model of polyelectrolyte complexation. Specifically, we built and
numerically simulated a field theory model of a salt-free solution
containing a symmetric mixture of flexible polyanions  and
polycations in a good solvent. This particular system constitutes
a minimal model for the phenomenon of complex coacervation, a type
of liquid-liquid phase separation in which a nearly pure solvent
phase coexists with second fluid phase (the ``coacervate'') that
contains the majority of the polyelectrolytes. Theoretically, the
symmetric coacervate model is interesting because the workhorse
tool of polymer physics, self-consistent field theory (SCFT),
fails to describe the electrostatic effects responsible for
coacervation.

Previous analytical work on closely related models of
polyelectrolyte mixtures has shown that complex coacervation can
be predicted based on calculations that assume weak, Gaussian
field fluctuations, i.e. calculations at the one-loop level of
fluctuation expansion~\cite{Borue:1988, Borue:1990,
Kudlay:2004a,Kudlay:2004b,Castelnovo:2001, Castelnovo:2002}.
However, the reliability of these predictions have remained
unclear because analytical techniques for treating more realistic
situations of strong charge and excluded volume correlations are
lacking. The FTS results of our paper are significant because they
can provide numerical data to assess the validity of the RPA, both
in terms of its predictions for charge and segment density
correlations and for the location of the two-phase envelope.
Overall, we have found that the RPA is remarkably robust in its
predictions, except at very high charge densities (large values of
the parameter $E$) where it overestimates the strength of charge
correlations and the size of the two-phase region.

Perhaps more significantly, our results have validated the
emerging field theoretic polymer simulation technique as a
powerful new tool for examining the structure and thermodynamics
of polyelectrolyte systems. FTS can be applied even in situations
where the RPA is inapplicable or technically very difficult, such
as cases of mesophases formed by weakly charged polymers with
hydrophobic backbones or block
co-polyelectrolytes~\cite{Yuri:2007}. Complexation of charged
polymers that also contain neutral blocks or grafts (which can be
either hydrophobic or hydrophilic) can also produce inhomogeneous
``structured coacervate'' phases that are not amenable to study by
current theoretical
methods~\cite{Har:1999,Bur:2004,Kramarenko:2003,Kramarenko:2006}.
We expect that FTS will prove to be a valuable tool for exploring
these and related types of polyelectrolyte systems.

\begin{acknowledgments}
The authors are grateful to Fyl Pincus and Kirill Katsov for many
valuable discussions and advice.  Acknowledgement is made to the
Donors of the American Chemical Society Petroleum Research Fund, the
Institute for Collaborative Biotechnology, Rhodia Corporation, and
the Mitsubishi Chemical Corporation for the support of this
research. This work made use of the MRL Computing Facilities
supported by the MRSEC Program of the National Science Foundation
under award No. DMR05-20415.
\end{acknowledgments}

\newpage
%\bibliography{fts}

\newpage
%figure captions
FIG.~1: {RPA structure factor for the segment density when
$BC=12$. Segment structure factors for several values of $N$,
calculated from Eq.~(\ref{eq:SNN_BCN}), are compared with the
result based on
the continuous Gaussian chain (CGC), cf. Eq.~(\ref{eq:SNN_BC}).}\\

FIG.~2: {The segment structure factor in solutions of neutral
discrete Gaussian chains: Symbols are FTS results calculated from
Eq.~(\ref{eq:SNN}). Lines are RPA structure factors from
FIG.~\ref{fig:SNN_RPA}. A three dimensional lattice of $M=32^3$
sites was employed with periodic boundary conditions. The cell
volume was $V=(4 R_g )^{3}$. $u_0$ was varied as $N$ was changed
to keep $B$ fixed at either $1$ or $12$. }\\

FIG.~3: {Inverse structure factor plot used to extract the
correlation length for segment density, $\xi_u$. According to
Eq.~(\ref{eq:xi_u}), $\xi_u$ can be estimated from the ratio of
the slope to intercept in the intermediate $k$ region of this
plot, $1/R_{g}< k/2\pi< 1/\xi_{u}$. We find that $\xi_{u}/R_g =
0.201\pm 0.0014$ in the case of $B=1$ and $C=12$, and $\xi_{u}/R_g
= 0.225\pm 0.0016$ for $B=12$ and $C=1$. The Edwards correlation
length is $\xi_{E}/R_g \simeq 0.204$ for $BC=12$. The simulations
were conducted on a $M=32^3$ lattice with periodic boundary
conditions, a system volume of $V=(4 R_g )^{3}$, and a chain
length of $N=1024$. The linear fit was applied in the regime of
${(kR_g)}^2\in[30, 200]$. }\\

FIG.~4: {The segment structure factor in solutions of symmetric
polyelectrolytes. The symbols are FTS results and the line is
calculated from the RPA result, Eq.~(\ref{eq:SNN_BCN}). The
following parameters were used: $B=1$, $C=12$, $N=1024$,
$V=4\times 4\times 4 {R_g}^{3}$, and $M=32\times 32\times 32$.}\\

FIG.~5: {The charge structure factor in solutions of oppositely
charged polyelectrolytes. Symbols are FTS results, while the lines
are RPA structure factors calculated from Eq.~(\ref{eq:Szz_CE}).
In the simulations, we vary the parameter $E$ by changing $z$
while keeping $l_B$ constant. The parameters correspond to:
$l_{B}=R_g/8$, $B=1$, $C=12$, $N=1024$, $V=(4 R_g )^3$, and
$M=32^3$.}\\

FIG.~6: {Phase diagram for the three-dimensional symmetric
polyelectrolyte mixture expressed in the coordinates of reduced
polymer concentration $C$ and reduced excluded volume $B$ at fixed
reduced Bjerrum length $E=14400$. Solid and dashed lines are the
analytical one-loop binodal and spinodal, respectively. The
one-phase region is above and to the right of the lines, and the
two-phase region is below and to the left. Symbols are from the
examination of the hysteresis in $\Delta{\rho}$ using FTS; details
of the procedure are explained in the text. FTS data obtained from
supercooling simulations (lowering $B$) are denoted by $\bullet$,
while superheating data are denoted by $\blacksquare$. The dotted
line is a power-law fit. The error in the numerical data is
comparable to the size of the symbols. Numerical simulations were
conducted in a cubic cell of size $(4 R_g )^3$ with periodic
boundary conditions and with $N=384$.}\\

\end{document}